\documentclass{emulateapj}
\usepackage{apjfonts}
\usepackage{color}

\newcommand{\pivec}{\mbox{\boldmath $\pi$}}
\newcommand{\muvec}{\mbox{\boldmath $\mu$}}

\lefthead{HAN ET AL.} 
\righthead{NEW INTERPRETATION OF OGLE-2013-BLG-0723}

\begin{document}
\title{A New Non-Planetary Interpretation of the Microlensing Event
OGLE-2013-BLG-0723}

\author{
Cheongho Han$^{1}$,
David P.~Bennett$^{2,3}$,
Andrzej Udalski$^{4}$,
and
Youn Kil Jung$^{1}$
}
\affil{$^{1}$Department of Physics, Institute for Astrophysics, Chungbuk National University, 371-763 Cheongju, Republic of Korea,\\
$^2$Code 667, NASA Goddard Space Flight Center, Greenbelt, MD 20771, USA \\
$^{3}$University of Notre Dame, Department of Physics, 225 Nieuwland Science Hall, Notre Dame, IN 46556-5670, USA,\\
$^{4}$Warsaw University Observatory, Al. Ujazdowskie 4, 00-478 Warszawa, Poland}


\begin{abstract}
Recently, the discovery of a Venus-mass planet orbiting a brown-dwarf host 
in a binary system was reported from the analysis of the microlensing event 
OGLE-2013-BLG-0723.  We reanalyze the event considering the possibility of 
other interpretations.  From this, we find a new solution where the lens is 
composed of 2 bodies in contrast to the 3-body solution of the previous 
analysis.  The new solution better explains the observed light curve than the 
previous solution with $\Delta\chi^2\sim 202$, suggesting that the new solution 
is a correct model for the event.   From the estimation of the physical parameters 
based on the new interpretation, we find that the lens system is composed of two 
low-mass stars with $\sim 0.2\ M_\odot$ and $\sim 0.1\ M_\odot$ and located at 
a distance $\sim 3$ kpc.  The fact that the physical parameters correspond to 
those of the most common lens population located at a distance with a large 
lensing probability further supports the likelihood of the new interpretation.  
Considering that two dramatically different solutions can approximately explain 
the observed light curve, the event suggests the need of carefully testing all 
possible lens-system geometries.
\end{abstract}


\section{Introduction}

Recently, the microlensing discovery of a Venus-mass planet (OGLE-2013-BLG-0723LBb)
 orbiting a brown dwarf in a binary system was reported by \citet{Udalski2015}.  
The discovery of the planet was of special scientific interest in many aspects.  
First, the planet itself is the lowest-mass planet among those discovered by using 
the microlensing method.  Second, the host of the planet is a substellar-mass brown 
dwarf suggesting the possibility that an ice-rock planets can be formed in the outer 
parts of the accretion disk around a brown dwarf.  Third, the planet belongs to a 
binary system where the planet orbits the lower-mass component of the binary. Finally, 
the planet/host mass ratio indicates that the system may be viewed either as a 
scaled-down version of a planet plus a star or as a scaled-up version of a moon plus 
a planet orbiting a star, suggesting that the formation processes of companions within 
accretion disks around stars, brown dwarfs, and planets are similar.

\begin{figure}[t]
\epsscale{1.1}
\plotone{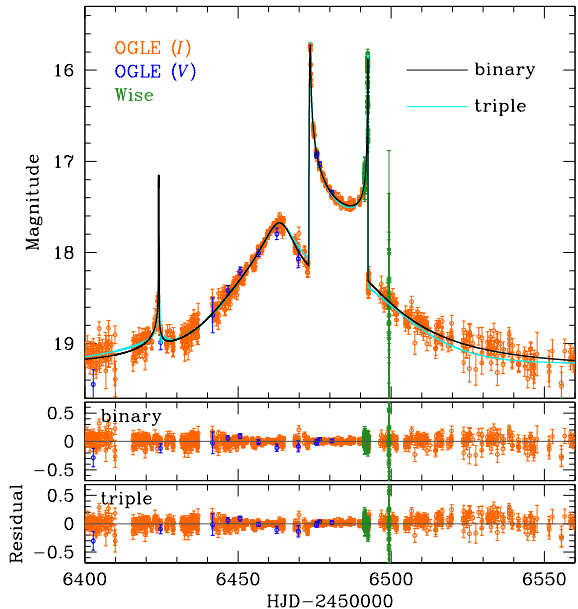}
\caption{\label{fig:one}
Light curve of OGLE-2013-BLG-0723. 
The cyan and black curves plotted 
over the data are the best-fit models obtained from the previous 3-body model and the 
newly found 2-body model, respectively. The two lower panels show the residuals from 
the individual models.
}
\end{figure}

The light curve of the microlensing event OGLE-2013-BLG-0723 is complex and composed 
of multiple anomalous features.  See the light curve in Figure~\ref{fig:one}.  In 
chronological order, there exists a short-term anomaly occurred at 
${\rm HJD}'={\rm HJD}-2450000\sim 6424$ followed by a bump at ${\rm HJD}'\sim 6463$ 
and two sharp spikes occurred at ${\rm HJD}'\sim 6472$ and 6492.  According to the 
interpretation of \citet{Udalski2015}, the main anomalous features (the bump and the 
two spikes) of the light curve were produced by a wide binary where the projected 
separation between the binary components is greater than the angular Einstein radius
$\theta_{\rm E}$ of the lens system and the short-term anomaly was produced by a 
planetary-mass object accompanied to the lower-mass component of the binary.  See the 
geometry of the lens system presented in Figure 2 of \citet{Udalski2015}.

For lensing light curves produced by binary objects, it is known that there can be a 
pair of degenerate solutions where one solution has a binary separation normalized to 
the Einstein radius, $s$, is greater than unity, $s>1$ (wide binary), and the other 
solution has a separation smaller than unity, $s<1$ (close binary) 
\citep{Griest1998, Dominik1999}.  According to the interpretation of \citet{Udalski2015}, 
the main features of the light curve were explained by a wide-binary solution.  They 
did find a close-binary solution\footnote{This solution is different from that one 
presented in this work.}, but the solution was rejected because it resulted in unphysical 
lens parameters.

In this work, we present another interpretation of the lensing event OGLE-2013-BLG-0723 
based on a new solution of lensing parameters found from the reanalysis of the event. 
According to this interpretation, all features of the lensing light curve including 
the short-term anomaly can be explained by a close-binary model, 
that was not found in the previous analysis,
without the need to introduce an additional planet.

\section{Reanalysis}

Microlensing light curves produced by binary objects are described by many parameters.  
For the simplest case of a rectilinear lens-source relative motion, one needs 7 basic 
parameters.  Among these parameters, 3 parameters describe the relative lens-source 
motion, including the time of the closest source approach to a reference position of 
the lens, $t_0$, the separation between the source and the reference position at $t_0$, 
$u_0$ (normalized to the angular Einstein radius $\theta_{\rm E}$), and the angle between 
the source trajectory and the binary axis, $\alpha$.  In our modeling, we use the barycenter 
of the binary as a reference position.  The Einstein time scale $t_{\rm E}$, which is 
defined as the time required for the source to cross the Einstein radius, is needed to 
characterize the time scale of an event. Another two parameters of the projected binary 
separation, $s$ (also normalized to $\theta_{\rm E}$), and the mass ratio between the 
binary components, $q$, characterize the binary lens components.  The last parameter is 
the normalized source radius $\rho=\theta_*/\theta_{\rm E}$, where $\theta_*$ is the 
angular source radius. This parameter is needed to describe the caustic-crossing features 
that are affected by finite-source effects. See Figure 6 of \citet{Jung2015} for the 
graphical presentation of the binary lensing parameters.

Modeling based on the basic parameters is often not enough to precisely describe lensing 
light curves and additional parameters are needed to consider higher-order effects. In 
order to consider parallax effects, which are caused by the positional change of the 
observer induced by the orbital motion of the Earth around the Sun \citep{Gould1992}, 
one needs two additional parameters $\pi_{{\rm E},N}$ and $\pi_{{\rm E},E}$.  These 
parameters represent the two components of the lens parallax vector $\pivec_{\rm E}$ 
projected onto the sky along the north and east equatorial coordinates, respectively. 
The magnitude of the parallax vector corresponds to $\pi_{\rm E}=\pi_{\rm rel}/\theta_{\rm E}$, 
where $\pi_{\rm rel}={\rm AU}(D_{\rm L}^{-1}-D_{\rm S}^{-1})$ is the relative lens-source 
parallax, $D_{\rm L}$ and $D_{\rm S}$ are the distances to the lens and source, respectively.
The direction of $\pivec_{\rm E}$ is that of the  relative lens-source motion in the frame 
of the Earth at a reference time of the event.

\begin{deluxetable*}{lrrrrrr}
\tablecaption{Lensing parameters\label{table:one}}
\tablewidth{0pt}
\tablehead{
\multicolumn{1}{c}{Parameter} & 
\multicolumn{1}{c}{Standard}  & 
\multicolumn{2}{c}{Parallax}  &
\multicolumn{1}{c}{Orbital}  & 
\multicolumn{2}{c}{Orbital + Parallax} \\
\multicolumn{1}{c}{} &
\multicolumn{1}{c}{} &
\multicolumn{1}{c}{$u_0>0$} &
\multicolumn{1}{c}{$u_0<0$} &
\multicolumn{1}{c}{} &
\multicolumn{1}{c}{$u_0>0$} &
\multicolumn{1}{c}{$u_0<0$} 
}
\startdata
$\chi^2$                       & 5536.9               &   4292.3               &    4211.1              &   3936.2                &   3925.1                &  3930.1                \\
$t_0$ (HJD')                   & $6485.057\pm 0.028$  &  $ 6485.055\pm 0.038$  &  $ 6485.285\pm 0.048$  &  $ 6485.976\pm 0.054$   &   $ 6486.004\pm 0.056$  &  $ 6486.007\pm 0.069$  \\   
$u_0$                          & $0.0133  \pm 0.001$  &  $ 0.019   \pm 0.002$  &  $-0.013   \pm 0.001$  &  $ 0.032   \pm 0.002$   &   $ 0.030   \pm 0.001$  &  $-0.028   \pm 0.001$  \\
$t_{\rm E}$ (days)             & $48.60   \pm 0.07 $  &  $ 55.24   \pm 0.22 $  &  $ 47.97   \pm 0.22 $  &  $ 51.93   \pm 0.42 $   &   $ 50.65   \pm 0.52 $  &  $ 51.33   \pm 0.48 $  \\
$s$                            & $0.668   \pm 0.001$  &  $ 0.670   \pm 0.001$  &  $ 0.665   \pm 0.001$  &  $ 0.672   \pm 0.003$   &   $ 0.672   \pm 0.003$  &  $ 0.666   \pm 0.003$  \\
$q$                            & $0.585   \pm 0.003$  &  $ 0.411   \pm 0.005$  &  $ 0.562   \pm 0.009$  &  $ 0.529   \pm 0.007$   &   $ 0.562   \pm 0.011$  &  $ 0.560   \pm 0.011$  \\
$\alpha$ (rad)                 & $4.892   \pm 0.002$  &  $-4.939   \pm 0.005$  &  $ 4.879   \pm 0.004$  &  $-4.991   \pm 0.005$   &   $-4.975   \pm 0.007$  &  $ 4.963   \pm 0.005$  \\
$\rho$ ($10^{-3}$)             & $1.04    \pm 0.01 $  &  $ 1.08    \pm 0.02 $  &  $ 1.06    \pm 0.02 $  &  $ 1.19    \pm 0.02 $   &   $ 1.19    \pm 0.02 $  &  $ 1.16    \pm 0.02 $  \\
$\pi_{{\rm E},N}$              & -                    &  $-1.14    \pm 0.04 $  &  $ 0.70    \pm 0.04 $  &  -                      &   $ 0.05    \pm 0.04 $  &  $ 0.10    \pm 0.07 $  \\
$\pi_{{\rm E},E}$              & -                    &  $ 0.04    \pm 0.05 $  &  $ 1.15    \pm 0.09 $  &  -                      &   $ 0.26    \pm 0.07 $  &  $ 0.30    \pm 0.08 $  \\
$ds/dt$ (${\rm yr}^{-1}$)      & -                    &    -                   &  -                     &  $ 0.06    \pm 0.04 $   &   $ 0.08    \pm 0.04 $  &  $-0.01    \pm 0.03 $  \\
$d\alpha/dt$ (${\rm yr}^{-1}$) & -                    &    -                   &  -                     &  $-0.46    \pm 0.03 $   &   $-0.53    \pm 0.03 $  &  $ 0.44    \pm 0.03 $ 
\enddata                                                                                            
\tablecomments{${\rm HJD}'={\rm HJD}-2450000$.}
\end{deluxetable*}

\begin{deluxetable}{lr}
\tablecaption{{\bf Lensing parameters of 3-body model}\label{table:two}}
\tablewidth{0pt}
\tablehead{
\multicolumn{1}{c}{Quantity} &
\multicolumn{1}{c}{Value}
}
\startdata
$\chi^2$                      &  4126.8 \\
$t_0$ (HJD')                  &  6484.526 $\pm$ 0.037 \\
$u_0$                         & -0.079    $\pm$ 0.002 \\
$t_{\rm E}$ (days)            &  68.48    $\pm$ 0.01 \\
$s_1$                         &  5.07     $\pm$ 0.02 \\
$q_1$                         &  3.11     $\pm$ 0.02 \\
$\alpha$ (rad)                & -1.195    $\pm$ 0.003 \\
$s_2$                         &  0.97     $\pm$ 0.02 \\
$q_2$ $(10^{-5})$             &  6.61     $\pm$ 0.01 \\
$\psi_0$ (rad)                & -4.936    $\pm$ 0.005 \\
$\rho$ $(10^{-3})$            &  1.40     $\pm$ 0.02 \\
$\pi_{{\rm E},N}$             & -0.05     $\pm$ 0.01 \\
$\pi_{{\rm E},E}$             &  1.35     $\pm$ 0.02 \\
$ds_2/dt$ $({\rm yr}^{-1})$   &  0.81     $\pm$ 0.02 \\ 
$d\psi/dt$ $({\rm yr}^{-1})$  & -0.50     $\pm$ 0.02  
\enddata  
\end{deluxetable}

Another higher-order effect that is often needed to consider in binary-lens modeling is the 
orbital motion of the lens \citep{Park2013}.  To first-order approximation, the lens-orbital 
effect is described by two parameters $ds/dt$ and $d\alpha/dt$, which represent the change 
rates of the binary separation and the source trajectory angle, respectively \citep{Albrow2000}.

Ideally, a solution of lensing parameters can be searched for by comparing an observed 
light curve with all possible model light curves resulting from the combination of 
lensing parameters. However, thorough coverage of the vast parameter space is limited 
by computing power and thus the grid spacing of each parameter cannot be arbitrarily 
small to completely cover the parameter space.  As a result, solutions of lensing parameters, 
especially ones located in a very localized region in the parameter space, can be occasionally 
missed.\footnote{An adaptive mesh refinement approach, which locally adds more grids
where they are needed, can help to minimize the probability of missing solutions, 
but such a code has not yet been developed in microlensing analysis due to the difficulty
in providing a prior condition for denser grid regions.}

The reanalysis of this event was initiated by one of us (DPB) using the initial condition
grid search method of \citet{Bennett2010} with the data set used in the original paper 
\citep{Udalski2015}. To limit the parameter space to be searched, the initial search was 
conducted using only a stellar binary model without microlensing parallax or orbital motion. 
This search yielded a candidate solution with a caustic crossing feature at 
${\rm HJD}'\approx 6414.5$ where there is a gap in the data. This is relatively close to 
the observed light curve bump at ${\rm HJD}'\approx 6424$, suggesting that a model including 
parallax and/or orbital motion might naturally explain the light curve. Several similar 
models were explored, and a model including microlensing parallax with the early light 
curve peak at ${\rm HJD}'\approx 6424$ was found. This solution was sent to the lead 
authors of the original paper, while models also including orbital motion were investigated.

Considering the possibility that there may exist missed solutions, especially in the close
binary regime, we conduct reanalysis of the lensing event OGLE-2013-BLG-0723.  In this 
analysis, we narrow down the grid spacing of parameters in order to minimize the possibility 
of missing localized solutions.  We also consider higher-order effects in the initial 
solution search to avoid the possibility that local solutions are missed due to the 
neglect of higher-order effects. The grid search is conducted in the space of the 
parameters $(s,q,\alpha)$ because lensing light curves can change dramatically with the 
small changes of these parameters.  To search for other parameters, for which lensing 
light curves vary smoothly with the change of the parameters, a downhill approach is used.
For the downhill approach, we use a Markov Chain Monte Carlo (MCMC) method.

It is needed to consider finite-source effects in order to describe the caustic-crossing 
parts of the light curve.  We incorporate the effects by using a numerical ray-shooting 
method.  In this method, uniform rays are shot from the image plane, bent by the lens 
equation, and then collected in the source plane.  Then, finite magnifications are computed 
as the ratio of the ray number density on the source plane to the density on the image plane.  
Since precise computations of finite magnifications require a large number of rays, this 
numerical method demands large amount of computing power.  For efficient production of 
theoretical light curves, we apply the ``map-making'' method \citep{Dong2006}, where one 
can produce many light curves resulting from different source trajectories based on a 
single ray-shooting map for a given set of the binary parameters $s$ and $q$.  In computing 
finite-source magnifications, we consider the surface-brightness variation of the source 
star by modeling the surface-brightness profile as 
$S_\lambda \propto 1-\Gamma_\lambda (3\cos\psi/2),$ where $\Gamma_\lambda$ is the linear 
limb-darkening coefficient $\psi$ is the angle between the line of sight toward the source 
center and the normal to the surface.  The limb-darkening coefficient $\Gamma_I=0.36$ is 
adopted from \citet{Claret2000} based on the source type.  For the detailed procedure of 
determining the source type, see section 4.

We note that it is difficult to consider lens-orbital effects in the initial search for 
solutions based on the  map-making method. This is because the binary separation and orientation 
vary in time during events and thus a single map cannot be used to produce multiple light curves.
We, therefore, consider orbital effects after a preliminary solution is found from the initial
search.

For direct comparison of models, we use the same data sets as those used in \citet{Udalski2015}. 
These data sets are composed of 4067 $I$-band and 19 $V$-band data acquired by the Optical 
Gravitational Lensing Experiment (OGLE) group and  62 $I$-band data obtained by the Wise group. 
We also use the same error-bar normalization.  Since $\chi^2$ per degree of freedom is normalized 
to unity, $\Delta\chi^2=n^2$ corresponds to the statistical importance of $n\sigma$ level.

\section{New Interpretation}

From the grid search, we find a new candidate close-binary solution that was not found in 
the previous analysis.  It turns out that the new solution was missed because the grid 
spacing was not small enough to locate the isolated solution in the parameter space.  
Once the approximate region of the solution in the parameter space is located, we gradually 
refine the solution first by allowing grid parameters to vary and then by considering 
higher-order effects.

Based on the newly found local minima, we test 5 different models. In the ``standard'' 
binary-lens model, we model the light curve based on the 7 basic lensing parameters.  
In the ``parallax'' and ``orbital'' models, we separately consider the lens parallax 
and orbital effects, respectively.  In the ``orbital + parallax'' model, we consider 
both lens-orbital motion and parallax effects.  In order to check the well-known 
``ecliptic degeneracy'' in the determination of the lens parallax \citep{Skowron2011}, 
we test two models with $u_0>0$ and $u_0<0$ for models considering parallax effects.

In Table~\ref{table:one}, we present the best-fit parameters of the individual tested 
models along with their $\chi^2$ values.  In Table~\ref{table:two}, we also present 
the parameters of the best-fit 3-body solution of \citet{Udalski2015} for comparison.  
We note that notations of the 3-body lensing parameters are different from those of the 
2-body parameters due to the addition of one more lens component.  For example, $(s_1,q_1)$ 
and $(s_2,q_2)$ denote the separations and mass ratios between $M_1-M_2$ and $M_1-M_3$ pairs, 
respectively, and $M_1$, $M_2$, and $M_3$ denote the lens components according to the order 
of heavier mass.  The notation $\psi_0$ denotes the angle between the binary axis and the 
planet-host axis at $t_0$ and $ds_2/dt$ and $d\psi/dt$ are the change rates of $s_2$ and 
$\psi$, respectively.

We find that higher-order effects are important for the precise description of the event.  
It is found that the standard model with $\chi^2=5536.9$ cannot explain the short-term 
anomaly. Furthermore, it leaves noticeable residuals in the regions of the light curve 
around the other anomalous features. Consideration of the higher-order effects significantly 
improves the fit. We find that inclusion of parallax effects improves the fit by 
$\Delta\chi^2=1326$. The improvement by considering lens-orbital effects is $\Delta\chi^2=1601$. 
When both higher-order effects are simultaneously considered, the improvement is $\Delta\chi^2=1612$. 
Considering that (1) $\chi^2$ improvement by the orbital effect is significantly greater than 
the improvement by the parallax effect and (2) the additional improvement from the orbital model 
to the orbital + parallax model is minor ($\Delta\chi^2=11$), we judge that the lens orbital 
motion is the dominant higher-order effect. With the inclusion of the higher-order effects, 
the lensing light curve fits all anomalous features.

\begin{figure}[t]
\epsscale{1.1}
\plotone{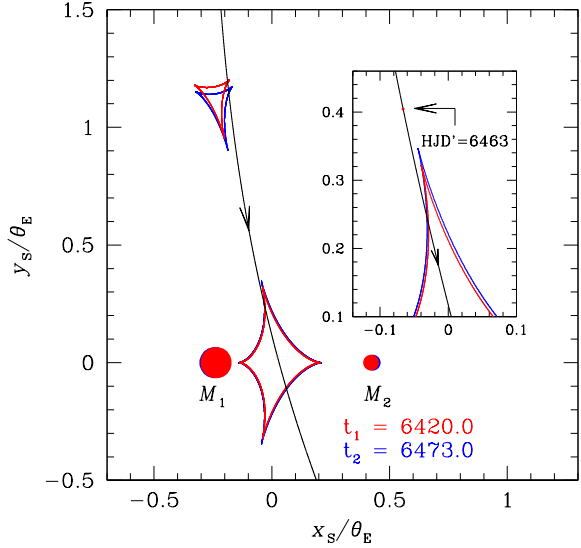}
\caption{\label{fig:two}
Geometry of the lens system.  The closed curves with cusps are the caustics of the 
lens and the curve with an arrow is the source trajectory. The two filled dots marked 
by $M_1$ and $M_2$ are the positions of the binary lens components.  All lengths are 
normalized to the angular Einstein radius $\theta_{\rm E}$ and the coordinates are 
centered at the barycenter of the binary lens.  The inset shows the zoom of the region 
around the upper sharp cusp of the central caustic.  The very tiny circle marked by 
${\rm HJD}'=6463$ represents the source position at the time of the bump located 
between the short-term anomaly and the caustic-crossing spikes in the lensing light 
curve. The size of the circle represents the source size scaled to the caustic size.  
Due to the orbital motion of the binary lens, the positions of the lens components 
and the corresponding caustics vary in time. We present the positions of the lens and 
caustics at two different epochs ${\rm HJD}'=6420$ and 6473, which correspond to the 
time of the short-term anomaly and the first source star's crossing of the central 
caustic, respectively. 
}
\end{figure}

\begin{figure*}[t]
\epsscale{0.9}
\plotone{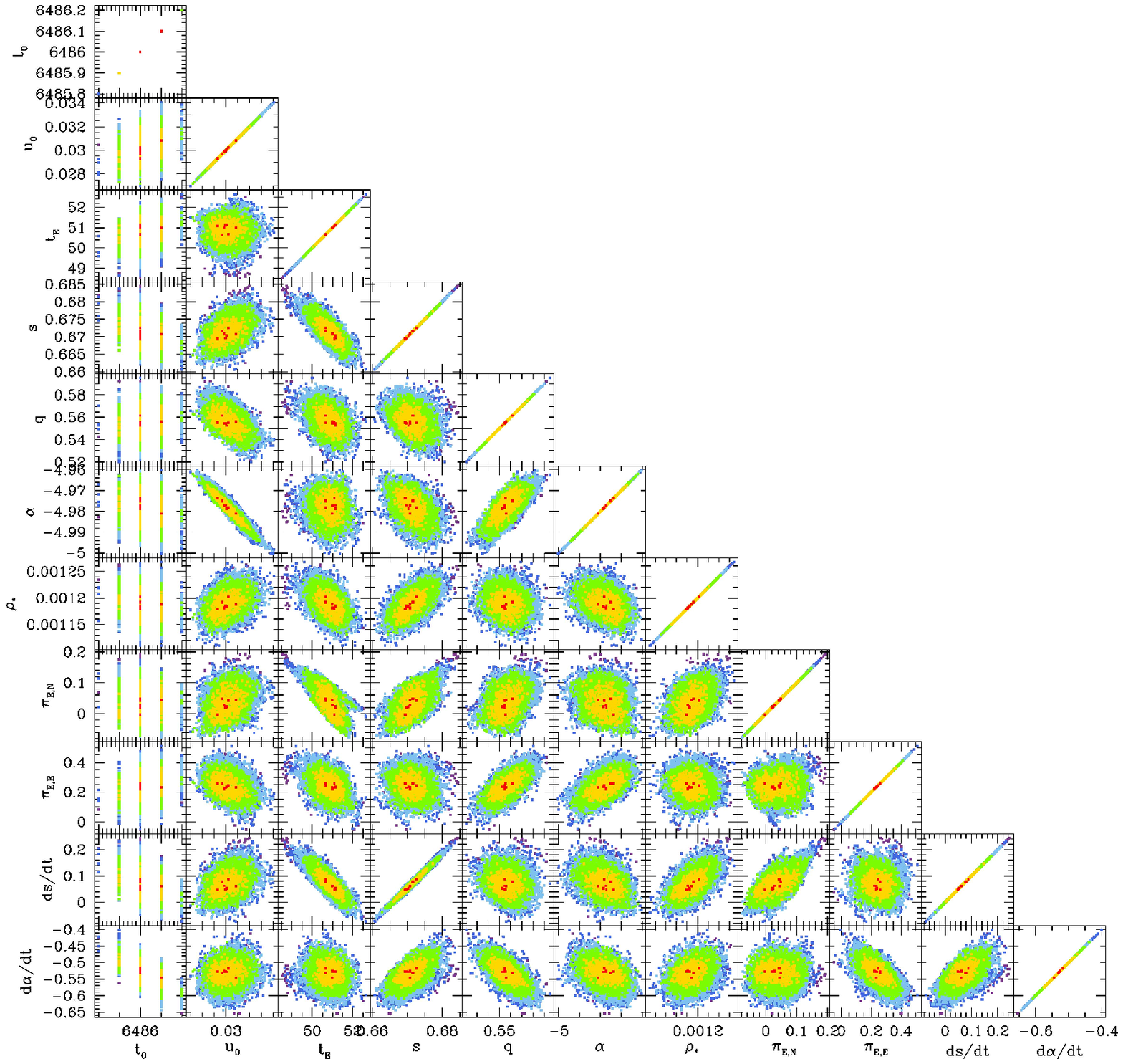}
\caption{\label{fig:three}
Distribution of the lensing parameters of the newly found binary solution. 
The color coding represents points on the MCMC chain within 
1 (red), 2 (yellow), 3 (green), 4 (cyan), 5 (blue) $\sigma$ of the best fit.
}
\end{figure*}

In Figure~\ref{fig:one}, we present the light curve of the close-binary model obtained 
from the reanalysis. For comparison, we also present the triple-lens model light curve 
that is constructed based on the best-fit parameters presented in \citet{Udalski2015}. 
Figure~\ref{fig:two} shows the lens-system geometry of the close binary model where 
the source trajectory with respect to the caustics are presented. According to the new 
close-binary model, the short-term anomaly was produced by the passage of the source 
over the tip of the 3-cusp peripheral caustic, and the two caustic-crossing spikes were 
produced by the source passage over the 4-cusp central caustic. The bump between the 
short-term anomaly and the caustic-crossing spikes was produced when the source passed 
over the narrow strip of the strong anomaly region formed along the line connecting the 
sharp cusps of the central and peripheral caustics.  In Figure~\ref{fig:three}, we also 
present the distribution of the lensing parameters of the best-fit solution.

\begin{figure}[t]
\epsscale{1.15}
\plotone{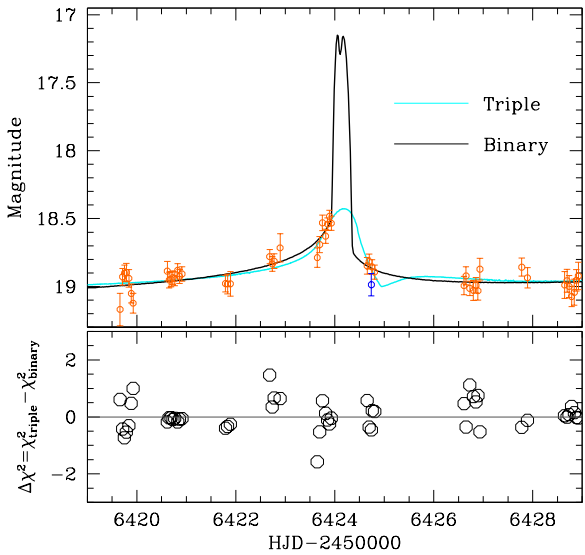}
\caption{\label{fig:four}
Enlargement of the light curve near the short-term anomaly around ${\rm HJD}'\sim 6424$. 
The lower panel shows the $\chi^2$ difference between the new binary and previous 
triple-lens solutions. Positive $\Delta\chi^2$ implies that the 2-body model 
provides a better fit and vice versa.  
}
\end{figure}

We find that both the previous 3-body and the new 2-body solutions almost equally well
explain the short-term anomaly.  In Figure~\ref{fig:four}, we present the enlargement 
of the anomaly region over which we plot both model light curves.  The lower panel shows 
the $\chi^2$ difference between the two models. Positive $\Delta\chi^2$ value means that 
the binary model provides a better fit and vice versa.  One finds that $\chi^2$ difference 
for nearly all data points are $\lesssim 1$, implying that the anomaly is well explained 
by both solutions.  We note that despite the almost the same goodness of the fits, the 
two models are greatly different; the binary solution predicts a huge 1.5 -- 2.0 mag 
brightening, while the triple model predicts a mild variation.  If there existed a few 
points at the peak of the anomaly, the two models could have been clearly distinguished.  
Unfortunately, this part of the light curve was not covered by data.

\begin{figure}[t]
\epsscale{1.1}
\plotone{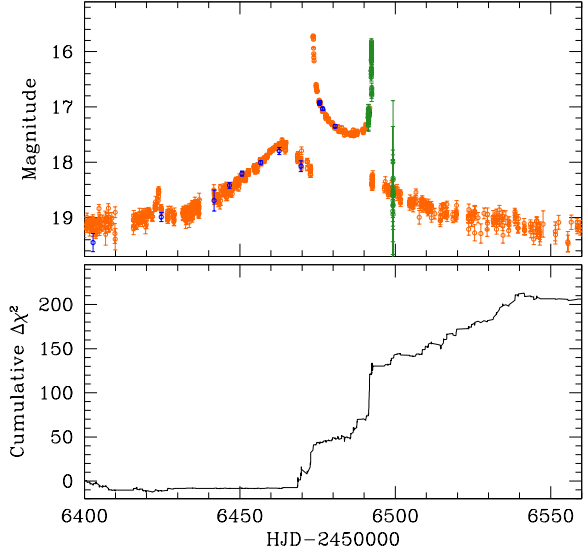}
\caption{\label{fig:five}
Cumulative distribution of $\Delta\chi^2$ between the 2-body and 3-body models
as a function of time. 
}
\end{figure}

Although it is difficult to resolve the degeneracy between the two solutions based on
the short anomaly, we find that the degeneracy can be resolved from the overall light 
curve.  One can see the goodness of the 2-body fit over the 3-body fit from the 
comparison of $\chi^2$ values of the fits presented in Tables~\ref{table:one} and 
\ref{table:two} and the cumulative function of $\Delta\chi^2$ between the two models 
as a function of time presented in Figure~\ref{fig:five}.  It is found that the 2-body 
solution with $\chi^2=3925.1$ provides a better fit over the 3-body solution with 
$\chi^2=4126.8$, i.e. $\Delta\chi^2=201.7$. From the cumulative distribution, it is 
found that the 2-body solution better explains the observed light curve in the regions 
between the caustic crossings and the declining part of the light curve after the caustic 
crossings.  We find that the $\chi^2$ difference between the two models is $\Delta\chi^2=202$.  
We note that $\chi^2\sim 50$ improvement is achieved during 
$6510 \lesssim {\rm HJD}'\lesssim 6540$, when the phase of the Moon was close to full and 
thus data show somewhat larger scatter.  We note, however, that the fit improvement is not 
attributed to the effect of elevated sky background or other noise because such an effect 
was accounted for in the error bar estimation during photometry procedure.  Furthermore, 
the sharp increases in the cumulative $\chi^2$ plot in other regions, especially around 
the caustic-crossing regions, demonstrate the better fit of the new solution.  
\citet{Udalski2015} mentioned a systematic trend in the baseline magnitude of the source 
star caused by a nearby bright star.  In Appendix, we show this trend.  We also 
describe the procedure to correct the systematic trend. We note that the improvement 
of the fit by the new solution is not attributed to this systematic trend because we use data 
set where the trend was corrected.  Considering that not only the new solution 
provides a better fit but also the 2-body model is simpler than the 3-body model, the 
2-body solution is likely to be the correct solution according to Occam's razor.

We consider two tests that may further support our new interpretation.  The first test 
is checking the brightnesses of the source and blend predicted by the two models from 
high-resolution observations.  We find that this method would not work because the 
estimated source and blend fluxes of the binary solution 
$(F_s,F_b)=(0.151\pm 0.001,0.170\pm 0.001)$ are not much different from 
$(F_s,F_b)=(0.191\pm 0.001,0.132\pm 0.001)$ of the triple-lens solution.  The second 
test is measuring the relative lens-source proper motion.  The detailed procedure of 
computing the proper motion is described in the next section.  We find that the 
heliocentric proper motion estimated from the binary model is 
$\mu_\odot = 6.4\pm 0.5\ {\rm mas\ yr}^{-1}$, while the expected value for the 
triple-lens model is $\mu_\odot = 14.5\pm 1.3\ {\rm mas\ yr}^{-1}$.  The difference 
between the estimated proper motions is considerable, implying that future observations 
with high-resolution instrument will be able to distinguish the two solutions.

In the aspect that there exist multiple interpretations for an isolated short-term anomaly 
located away from the main light curve, OGLE-2013-BLG-0723 is similar to the case of 
MACHO-97-BLG-41.  For MACHO-97-BLG-41, there existed two interpretations: one interpretation 
by a 3-body circumbinary planetary system \citep{Bennett1999} and the other by a 2-body 
orbiting binary system \citep{Albrow2000, Jung2013}. Considering that two solutions with 
dramatically different lens systems can explain observed light curves, the two events 
suggest the need of carefully testing all possible lens-system geometries.

\section{New Estimation of Physical Parameters}

\begin{figure}[t]
\epsscale{1.1}
\plotone{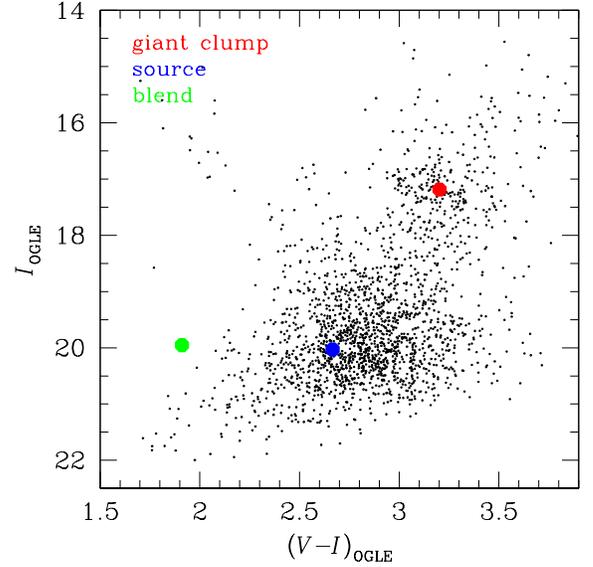}
\caption{\label{fig:six}
Positions of the source star and the blend with respect to the centroid 
of the giant clump in the color-magnitude diagram.  The magnitude and color 
are calibrated to the OGLE-IV photometry.  
}
\end{figure}

For the new solution, both finite-source and parallax effects are detected and thus we 
are able to determine the angular Einstein radius $\theta_{\rm E}$ and the lens parallax 
$\pi_{\rm E}$. With these values, the mass and distance to the lens are uniquely determined by 
\begin{equation}
M_{\rm tot}={\theta_{\rm E} \over \kappa \pi_{\rm E}};\qquad
D_{\rm L}={{\rm AU} \over \pi_{\rm E}\theta_{\rm E}+\pi_{\rm S}},
\end{equation}
where $\kappa=4G/(c^2{\rm AU})$ and $\pi_{\rm S}={\rm AU}/D_{\rm S}$ is the parallax of 
the source star \citep{Gould2000}. The lens parallax is determined from modeling. The 
angular Einstein radius is estimated from the relation $\theta_{\rm E}=\theta_*/\rho$, 
where the normalized source radius $\rho$ is measured from modeling and the angular source 
radius $\theta_*$ is estimated from the source type that is determined based on the 
de-reddened color and brightness. For the calibration of the color and brightness of the 
source star, we use the centroid of bulge giant clump as a reference \citep{Yoo2004}. 
In Figure~\ref{fig:six}, we present the position of the source with respect to the 
centroid of the giant clump in the color-magnitude diagram.  The 
estimated de-reddened color and brightness of the source star are $(V-I,I)_0=(0.52\pm 0.05,17.3\pm 0.01)$, 
indicating that the source is an F-type main-sequence star. We then convert $V-I$ color 
into $V-K$ color using the color-color relation of \citet{Bessell1988} and obtain the angular 
source radius $\theta_*$ using the relation between $V-K$ and $\theta_*$ presented in 
\citet{Kervella2004}. The estimated source radius is $\theta_*=0.90 \pm 0.06\ \mu{\rm as}$. 
We note that the source radius is slightly smaller than the value estimated in 
\citet{Udalski2015} because of the slight color difference, which is caused by the difference 
in the ratios of the source to blended light between the two models.  From the angular source 
radius, the estimated the angular Einstein radius is $\theta_{\rm E}=0.75 \pm 0.05$ mas.

\begin{deluxetable*}{lrr}
\tablecaption{Physical Lens parameters\label{table:three}}
\tablewidth{0pt}
\tablehead{
\multicolumn{1}{c}{Parameter} & 
\multicolumn{1}{c}{$u_0>0$}   & 
\multicolumn{1}{c}{$u_0<0$} 
}
\startdata
Primary mass                                       &  $0.22 \pm 0.06\ M_\odot$              &   $0.19 \pm 0.04\ M_\odot$             \\
Companion mass                                     &  $0.13 \pm 0.04\ M_\odot$              &   $0.10 \pm 0.02\ M_\odot$             \\   
Distance to the lens                               &  $3.11 \pm 0.57$ kpc                   &   $2.74 \pm 0.45$ kpc                  \\
Projected separation                               &  $1.57 \pm 0.29$ AU                    &   $1.38 \pm 0.23$ AU                   \\
Geocentric lens-source proper motion               &  $5.42 \pm 0.39$ ${\rm mas\ yr}^{-1}$  &   $5.39 \pm 0.39$ ${\rm mas\ yr}^{-1}$ \\
Heliocentric lens-source proper motion (north)     &  $0.91 \pm 0.07$ ${\rm mas\ yr}^{-1}$  &   $1.58 \pm 0.11$ ${\rm mas\ yr}^{-1}$ \\ 
Heliocentric lens-source proper motion (east)      &  $6.43 \pm 0.47$ ${\rm mas\ yr}^{-1}$  &   $6.46 \pm 0.46$ ${\rm mas\ yr}^{-1}$ \\
Ratio of the projected kinetic to potential energy &  0.04                                  &   0.02                                  
\enddata                                                                                            
\end{deluxetable*}

In Table~\ref{table:three}, we list the determined physical parameters.  Although the $u_0>0$ 
solution is slightly preferred over the $u_0<0$ solution, it is found that the $\chi^2$ 
difference between the two models is merely $\Delta\chi^2=5.0$.  Since such a level of  
$\Delta\chi^2$ can be ascribed to systematics in data, we present the physical parameters 
corresponding to both solutions. We note, however, that the estimated lens parallax values 
from the two models are similar each other and thus the resulting physical parameters are 
also similar. Once the distance to the lens is known, the projected separation between the 
binary components and the geocentric lens-source proper motion are estimated by
\begin{equation}
a_\perp=s D_{\rm L}\theta_{\rm E}
\end{equation}
and
\begin{equation}
\mu_\oplus={\theta_{\rm E}\over t_{\rm E}},
\end{equation}
respectively.  We note that the geocentric reference frame is set with respect to the Earth 
position at $t_0$ (${\rm HJD}\sim 2456486$), which approximately corresponds to the time of 
the source star's closest approach to the center of mass of the binary lens.  The heliocentric 
proper motion is estimated from the geocentric proper motion by 
\begin{equation}
\muvec_\odot = \mu_\oplus {\pivec_{\rm E}\over \pi_{\rm E}} + 
{{\bf v}_{\oplus,\perp} \over {\rm AU}} \pi_{\rm rel},
\end{equation}
where ${\bf v}_{\oplus,\perp}=(-2.1,26.3)\ {\rm km\ s}^{-1}$ is the velocity of the Earth 
projected on the sky at the reference time, i.e. $t_0$.
We also present ratio of the projected kinetic to potential energy \citep{Dong2009} 
that is computed by
\begin{equation}
\left({{\rm KE}\over{\rm PE}}\right)_\perp =
{ (r_\perp/{\rm AU})^2 \over 8\pi^2(M_{\rm tot}/M_\odot)}
\left[ 
\left( {1\over s}{ds\over dt}\right)^2  + 
\left( {d\alpha\over dt}\right)^2
\right].
\end{equation}
To be a bound system, 
the ratio of the binary lens should follow $({\rm KE}/{\rm PE})_\perp \leq {\rm KE}/{\rm PE} \leq 1$.
For both the $u_0>0$ and $u_0<0$ solutions, this condition is satisfied.

The characteristics of the lens determined in the new analysis is greatly different from 
those determined in the previous analysis. The newly estimated masses of the binary 
components are $M_1\sim 0.2\ M_\odot$ and $M_2\sim 0.1\ M_\odot$, indicating that they 
are low-mass stars, which are the most common population of Galactic lenses. On the other 
hand, the previous analysis characterized the lens system as a 3-body system where a 
terrestrial planet is orbiting sub-stellar brown-dwarf host accompanied by a low-mass 
companion. The distance to the lens estimated in the new analysis is $D_{\rm L}\sim 3$ kpc, 
which corresponds to roughly a halfway distance between the observer and the source and 
thus the lensing probability is maximized.  By contrast, the previous analysis estimated 
a very close lens distance of $D_{\rm L}\lesssim 500$ for which the lensing probability 
is low.  Therefore, the likelihood of the new interpretation is further supported by the 
fact that the physical parameters of the lens system correspond to the most probable 
values predicted by the mass function and distribution of Galactic matter.

\section{Summary}

We presented a new interpretation of the microlensing event OGLE-2013-BLG-0723 that 
had been previously interpreted as a 3-body lensing event produced by a Venus-mass 
planet orbiting a brown-dwarf host in a binary system.  The new solution, where the 
lens is composed of 2 bodies, can explain all anomalous features in the lensing light 
curve without the need to introduce an additional planetary companion.  The fact that 
the new solution better explains the observed light curve than the previous solution 
with a simpler model implies that the new model is preferred, a conclusion that is 
also supported by Occam's razor.  In addition, the fact that the physical parameters 
estimated from the new solution correspond to those of the most common lens population 
located at a distance of a large lensing probability further supports the likelihood 
of the new interpretation.  Considering that two dramatically different solutions can 
explain the observed light curve, the event suggests the need of carefully testing all 
possible lens-system geometries.  In particular, care must be taken to ensure that 
close binary solutions with caustic motion due to orbital motion are considered. 
OGLE-2013-BLG-0723 is the second close stellar binary microlensing event that was 
originally misinterpreted as a binary star system with a planet.

\acknowledgments
Work by C.~Han was supported by Creative Research Initiative Program
(2009-0081561) of National Research Foundation of Korea. 
D.P.B. was supported by grants NASA-NNX13AF64G and NNX15AJ76G.
The OGLE project has received funding from the National Science Centre,
Poland, grant MAESTRO 2014/14/A/ST9/00121 to AU.

\appendix

\begin{figure}[t]
\epsscale{0.7}
\plotone{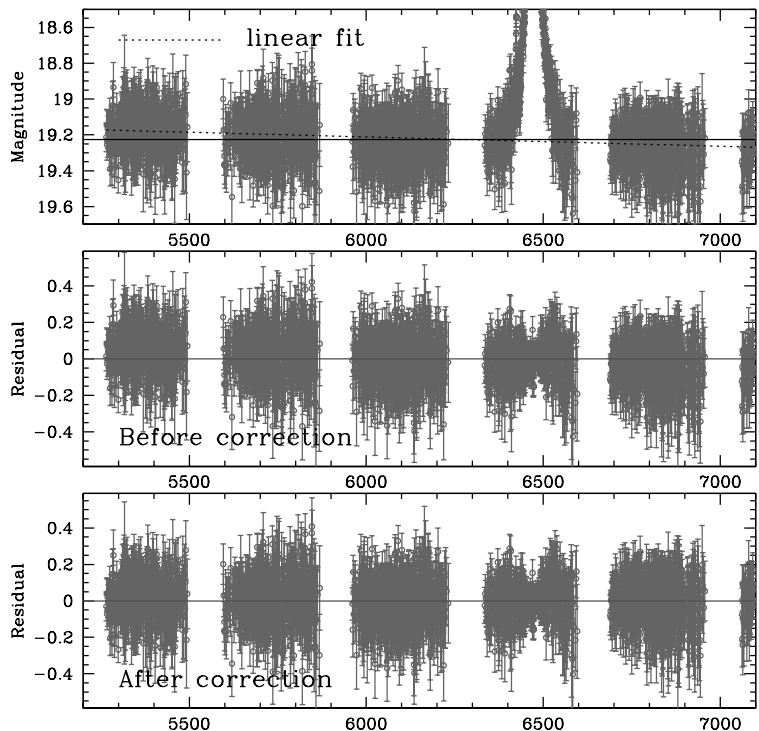}
\caption{\label{fig:seven}
Systematic trend in the baseline flux (uper panel) and the procedure to 
correct the trend.}
\end{figure}

\section{Correction of Baseline Variation}

The light curve of OGLE-2013-BLG-0723 shows a systematic decline in the baseline. See
the upper panel of Figure~\ref{fig:seven}. A similar long term linear trend (of opposite sign) was seen in
OGLE-2013-BLG-0341 and was eventually traced to a nearby bright star that was gradually
moving toward (in that case) the source star, so that more of its flux was being "captured" in
the tapered aperture used to estimate the source flux. We searched for such a moving bright
star by examining the difference of two images, from 2004 and 2012. We indeed find a dipole
from a bright star roughly 1.5" away, which is the characteristic signature of such moving stars.
Having identified the cause of this trend, we conduct a linear fit for it and remove it.
In the middle and lower panels of Figure~\ref{fig:seven}, we present the baseline 
of the source star before and after the correction, respectively.


\begin{thebibliography}{99}

\bibitem[Albrow et al.(2000)]{Albrow2000} Albrow, M.~D., Beaulieu, J.-P., Caldwell, J.~A.~R., et al. 2000, \apj, 534, 894
\bibitem[Bennett(2010)]{Bennett2010} Bennett, D.P.\ 2010, \apj, 716, 1408
\bibitem[Bennett et al.(1999)]{Bennett1999} Bennett, D. P., Rhie, S. H., Becker, A. C., et al. 1999, Natur, 402, 57
\bibitem[Bessell \& Brett(1988)]{Bessell1988} Bessell, M.~S., \& Brett, J.~M. 1988, \pasp, 100, 1134
\bibitem[Claret(2000)]{Claret2000} Claret, A.\ 2000, \aap, 363, 1081
\bibitem[Dominik(1999)]{Dominik1999} Dominik, M.\ 1999, \aap, 341, 943
\bibitem[Dong et al.(2009)]{Dong2009} Dong, S., Gould, A., Udalski, A., et al. 2009, \apj, 695, 970
\bibitem[Dong et al.(2006)]{Dong2006} Dong, Subo, DePoy, D.~L., Gaudi, B. S., et al.\  2006, \apj, 642, 842
\bibitem[Gould(1992)]{Gould1992} Gould, A. 1992, \apj, 392, 442
\bibitem[Gould(2000)]{Gould2000} Gould, A. 2000, \apj, 542, 785
\bibitem[Griest \& Safizadeh(1998)]{Griest1998} Griest, K., \& Safizadeh, N.\ 1998, \apj, 500, 37
\bibitem[Jung et al.(2013)]{Jung2013} Jung, Y. K., Han, C., Gould, A., \& Maoz, D. 2013, \apj, 768, L7
\bibitem[Jung et al.(2015)]{Jung2015} Jung, Y. K.,Udalski, A., Sumi, T., et al.\ 2015, \apj, 798, 123
\bibitem[Kervella et al.(2004)]{Kervella2004} Kervella, P., Bersier, D., Mourard, D., et al. 2004, \aap, 428, 587
\bibitem[Park et al.(2013)]{Park2013} Park, H., Udalski, A., Han, C., et al.\ 2013, \apj, 778, 134
\bibitem[Skowron et al.(2011)]{Skowron2011} Skowron, J., Udalski, A., Gould, A., et al. 2011, \apj, 738, 87
\bibitem[Udalski et al.(2015)]{Udalski2015} Udalski, A., Jung, Y.~K., Han, C., et al.\ 2015, \apj, 812, 47
\bibitem[Yoo et al.(2004)]{Yoo2004} Yoo, J., DePoy, D.~L., Gal-Yam, A., et al. 2004, \apj, 603, 139

\end{thebibliography}
\end{document}